\documentclass[prl,twocolumn,superscriptaddress]{revtex4}

\usepackage{amsfonts}
\usepackage{graphicx,graphics,epsfig,times,bm,bbm,amssymb,amsmath,amsfonts,mathrsfs}
\usepackage[normalem]{ulem}
\usepackage{wrapfig}
\usepackage{setspace}
\usepackage{subfigure}
\usepackage{dsfont}
\usepackage{braket}
\usepackage[pdftex]{color}
\usepackage[pdfstartview=FitH]{hyperref}
\usepackage{upgreek }
\usepackage[x11names]{xcolor}
\usepackage{tikz}
\usepackage{natbib}
\usepackage{chngcntr}

\newenvironment{proof}[1][Proof]{\noindent\textbf{#1.} }{\ \rule{0.5em}{0.5em}}

\newtheorem{mylemma}{Lemma}

\newtheorem{mytheorem}{Theorem}

\newcommand{\bes} {\begin{subequations}}
\newcommand{\ees} {\end{subequations}}
\newcommand{\bea} {\begin{eqnarray}}
\newcommand{\eea} {\end{eqnarray}}
\newcommand{\beq} {\begin{equation}}
\newcommand{\eeq} {\end{equation}}

\def\>{\rangle}
\def\<{\langle}
\def\Tr{\textrm{Tr}}

\newcommand{\ketbra}[2]{|{#1}\>\!\<#2|}
\newcommand{\bracket}[2]{\<{#1}|{#2}\>}

\newcommand{\eps}{\varepsilon}

\newcommand{\ignore}[1]{}

\def\Eg{E_{\textrm{gap}}}
\def\mC{\mathcal{C}}
\def\PC{P_{\mathcal{C}}}

\begin{document}

\title{Quantum error suppression with commuting Hamiltonians: Two-local is too local}
\author{Iman Marvian}
\affiliation{Center for Quantum Information Science and \& Technology}
\affiliation{Department of Physics}
\author{Daniel A. Lidar}
\affiliation{Center for Quantum Information Science and \& Technology}
\affiliation{Department of Physics}
\affiliation{Department of Electrical Engineering}
\affiliation{Department of Chemistry, University of Southern California, Los Angeles, California 90089, USA}

\begin{abstract}
We consider error suppression schemes in which quantum information is encoded into the ground subspace of a Hamiltonian comprising a sum of commuting terms. Since such Hamiltonians are gapped they are considered natural candidates for protection of quantum information and topological or adiabatic quantum computation. However, we prove that they cannot be used to this end in the $2$-local case. By making the favorable assumption that the gap is infinite we show that single-site perturbations can generate a degeneracy splitting in the ground subspace of this type of Hamiltonians which is of the same order as the magnitude of the perturbation, and is independent of the number of interacting sites
and their Hilbert space dimensions,
just as in the absence of the protecting Hamiltonian. This splitting results in decoherence of the ground subspace, and we demonstrate that for natural noise models the coherence time is  proportional to the inverse of the degeneracy splitting. Our proof involves a new version of the no-hiding theorem which shows that quantum information cannot be 
approximately hidden in the correlations between two quantum systems, and should be of independent interest. The main reason that $2$-local commuting Hamiltonians cannot be used for quantum error suppression is that their ground subspaces have only short-range (two-body) entanglement.
\end{abstract}

\maketitle

\textit{Introduction}.---%
Storing quantum information in the degenerate ground subspace $\mC$ of a Hamiltonian $H_0$, protected by an energy gap $\Eg$, is an appealing and ubiquitous idea with applications in topological \cite{Kitaev:97}, holonomic \cite{HQC}, and adiabatic quantum computation \cite{PhysRevA.74.052322}. Conventional wisdom states that the greater the gap, the better the protection, and that an infinite gap should provide absolute protection. This idea is based primarily on protection against thermal excitations because if the Hamiltonian is gapped, then thermal excitations are suppressed by a Boltzmann factor $e^{-\Eg/k_B T}$.
Thus, thermal excitations are completely suppressed at any finite temperature $T$ if $\Eg \rightarrow \infty$. However, decoherence is still a problem even in the absence of thermal excitations, i.e., when $T=0$. Here too an infinite gap is useful because it effectively transforms a perturbation $V$ added to $H_0$ into $V'=\PC V \PC$, where $\PC$ is the projector onto $\mC$, and $V'$ may be less damaging than $V$. In particular, one can choose $\mC$ such that $V'$ does no damage to the encoded quantum information for all local perturbations $V$. To have this property it is crucial that every state in the ground subspace $\mC$ is sufficiently entangled, since otherwise a local error can always break the ground subspace degeneracy. Thus \emph{both} a large gap and entanglement are needed for robust storage of quantum information in a degenerate ground subspace.
Here we shall assume that the large gap condition is satisfied and focus on the role of entanglement in providing a robust quantum memory. To this end we shall consider the important case of Hamiltonians $H_0$ that are sums of commuting terms, since they are automatically gapped \cite{Bravyi03}.  This type of Hamiltonians and the codes associated with them have been the subject of many recent studies \cite{Bravyi03, BPT_Com, Ahar11, Hastings_Top, Haah_Preskill}. They are a natural generalization of stabilizer codes \cite{Gottesman:1996fk}, which have been used in a variety of different Hamiltonian error suppression settings, in particular for topological quantum computing \cite{Kitaev:97,Dennis:02,Bombin:2014kx}, self-correcting quantum memories \cite{Bacon:05,Chesi:2010rc,Haah:2011fv,Hastings:2014uq} and adiabatic quantum computing \cite{PhysRevA.74.052322,PhysRevLett.100.160506,Young:13,Ganti:2014db,Mizel:2014dp}. We are particularly interested in the case of two-body interactions since they are easy to realize experimentally. Informally, we shall prove that \emph{codes defined as the ground subspaces of Hamiltonians with at most two-body interactions are not useful for quantum error suppression}. To formalize this we define, as usual, a $k$-local term as acting on at most $k$ sites, where each site $i$ is a $d_i$-dimensional quantum system (qudit), with $d_i \geq 2$. 
Our central result is the following no-go theorem:
\begin{mytheorem}
\label{Thm:main}
For any code defined as the degenerate ground subspace of a Hamiltonian $H_0$ that is a sum of 
$2$-local commuting terms, irrespective of the number of qudits or their Hilbert space dimensions, there exists a single-site perturbation which induces a degeneracy splitting that is equal to the magnitude of the perturbation up to a constant of order one, just as in the unprotected case where $H_0=0$. This results in a coherence time that is upper bounded by the inverse magnitude of the perturbation, just as the coherence time in the unprotected case. 
\end{mytheorem}
The implication is, clearly, that there is no advantage to using such codes, since they are no more effective than doing nothing. The culprit is the assumption of a $2$-local $H_0$; it turns out that its ground subspace is insufficiently entangled, and so even an infinite gap and arbitrarily high-dimensional qudits do not help. This may seem surprising, 
since there exist codes that exploit an infinitely large $d_i$ \cite{Got2}, and also since the larger is $d_i$ the smaller is the set of perturbations which are sums of local terms relative to the set of all perturbations, i.e., the stronger is the constraint of locality of perturbations, and this would appear to be advantageous for suppressing local perturbations.         
Our no-go theorem joins other results delimiting the possibilities for passive quantum information storage \cite{Bravyi:2009rz,Pastawski:2009fk,Yoshida:2011fk}. We set the bar even higher by showing that the simplest possible error model (single-site perturbations) is already sufficient to dash the hope of a passive, stable quantum memory built on physically reasonable two-body interactions, even under the most favorable assumption of an infinite $\Eg$.  
A finite $\Eg$ or a more elaborate noise model will only make matters worse. Circumventing the no-go theorem requires either $k$-local Hamiltonians with $k>2$ (such as in the toric code \cite{Kitaev:97}), which are difficult to realize experimentally; a non-commutative setting (such as might arise when using subsystem \cite{Bacon:05,Kribs:2005:180501}); or the addition of additional ingredients such as active quantum error correction \cite{Lidar-Brun:book}. 

Theorem~\ref{Thm:main} implies that the ground subspace of commuting $2$-local Hamiltonians cannot have topological order. In the special case of $1$-dimensional systems, if the interactions are also geometrically local we can find a stronger result. In these systems by blocking $k-1$ sites into one site, every $k$-local Hamiltonian can be mapped to a $2$-local Hamiltonian. Therefore,  Theorem~\ref{Thm:main} implies that in the one-dimensional systems the ground subspace of a geometrically local Hamiltonian which is sum of $k$-local commuting terms cannot have topological order.

To prove Theorem~\ref{Thm:main} we proceed in two parts. 
In the first part we introduce the ``induced degeneracy splitting" (IDS), a quantity that determines how much the degeneracy of the code subspace $\mC$ is split due to the perturbation $V$. We demonstrate that the IDS quantifies the deviation from the performance of an ideal quantum error detection code, and sets the decoherence timescale associated with the perturbation: the larger the IDS, the shorter the coherence time of the protected ground subspace. To demonstrate the latter we consider a simple  but general model of a perturbation with an unknown magnitude.
In the second part we restrict the perturbation to single sites and derive bounds on the IDS in the setting of two-body commuting Hamiltonians. 
To do so we first prove a new version of the ``no-hiding theorem" \cite{Cleve:99a,Braunstein:2007th,Kretschmann:2008zp}. 
In the following, for the sake of conciseness of presentation we present most of our results without proofs. Omitted details and proofs of all our technical results are given in the Appendix.

\textit{Effect of a perturbation in the large gap limit}.---%
Consider a gapped Hamiltonian $H_0$ and 
let $\mC$ be the $d$-dimensional ground subspace of $H_{0}$. We use the ``code" $\mC$ to encode protected quantum information. 
Next consider a perturbation $V_\lambda$ where $\lambda$ is a noise parameter,
so that the total Hamiltonian acting on the system is $H_\lambda = H_{0}+V_\lambda$.  
The perturbation $V_{\lambda}$ may describe imperfections in realizing the desired Hamiltonian $H_{0}$, or it may describe unknown local fields. In general a state in $\mC$ need not be an eigenstate of $H_\lambda$, and $V_{\lambda}$ will cause states which are initially in $\mC$ to acquire relative phases or even evolve outside of $\mC$. If we know the exact perturbation $V_{\lambda}$ and keep track of time then in principle we can always recover the initial state by applying a unitary transformation which undoes this time evolution.
However, usually the exact perturbation $V_{\lambda}$ is not known and our knowledge about $V_\lambda$ is described by a probability distribution $p(\lambda)$. 
Hence after a time $t$ the system evolves from an initial state $\rho(0)$ via a random unitary channel \cite{Audenaert:08} to 
\beq
\rho(t) = \int d\lambda\ p(\lambda) U_{\lambda}(t) \rho(0) U^{\dag}_{\lambda}(t)\ ,\quad U_{\lambda}(t)=e^{-i t H_{\lambda}}
\label{eq:rho}
\eeq 
(we use units where $\hbar=1$ throughout). This means that  uncertainty about the exact Hamiltonian of the system will lead to decoherence and loss of quantum information. We note that Eq.~\eqref{eq:rho} can also be derived from an open quantum system model, where the system-bath interaction $H_{I}=\sum_\lambda V_\lambda\otimes \ketbra{\lambda}{\lambda}$, with $\{\ket{\lambda}\}$ an orthonormal basis for the bath Hilbert space, and where both the bath Hamiltonian and the initial bath state $\rho_B$ commute with each $\ketbra{\lambda}{\lambda}$; then $p(\lambda)=\bra{\lambda}\rho_B\ket{\lambda}$.

One might expect that to suppress the effect of the noise it suffices to use a stronger protecting Hamiltonian $H_{0}$. This works to suppress thermal excitations, which vanish in the limit where the ground state gap $\Eg \gg kT$. But, it turns out that even in the limit where $\Eg \gg \|V_{\lambda}\|$ (where henceforth we use $\|\cdot \|$ to denote the operator norm, i.e., the largest singular value), the effect of $V_{\lambda}$ does not completely vanish on the code subspace. This is summarized in the following lemma \cite{ML:14} (see also the recent Refs.~\cite{Zanardi:2014fr,Bookatz:2014uq} for related results):
\begin{mylemma}\label{lemma_pert}
Let $\PC$ be the projector onto the zero-energy ground subspace of $H_0$ and \emph{$\Eg$} be the energy gap between the ground state and the first excited state. Then, for any perturbation $V_\lambda$:
\begin{equation}
\|e^{-i t(H_0+V_\lambda)} \PC -e^{-i t \PC V_\lambda \PC} \PC\|\le \frac{4\|V_\lambda\|}{\emph{$\Eg$}}(\|V_\lambda\| |t|+1) \ .
\label{lemma-pert}
\end{equation} 
\end{mylemma}
It follows that the effect of all $V_{\lambda}$-induced excitations vanishes when $\Eg\rightarrow\infty$, and in particular that in this limit 
\beq
U_\lambda(t) \PC {\rightarrow} \exp(-it V'_\lambda) \PC \ , \quad V'_\lambda = \PC  V_\lambda \PC \ .
\label{eq:Heff}
\eeq 
Thus, the perturbation acts on states in ${\mC}$ as the effective Hamiltonian $V'_\lambda$. The same effective Hamiltonian is the effect of $V_\lambda$ on the code subspace in standard first order degenerate perturbation theory \cite{Suzuki:1983cs}.

From these observations we find that in the large gap limit the code subspace remains invariant if and only if 
\begin{equation}
\label{eq:perfect}
\forall V_{\lambda}:\ \ \   \PC  V_{\lambda} \PC = \alpha_{\lambda} \PC 
\end{equation}
where $\alpha_{\lambda} $ is a constant which in general depends on $V_\lambda$. This condition means that in this limit states in $\mC$ do not evolve apart from a global phase, or equivalently, that their degeneracy is preserved. On the other hand, if Eq.~\eqref{eq:perfect} is not satisfied then the noise has a non-trivial effect on $\mC$ which does not vanish even in the large gap limit.

\textit{Induced degeneracy splitting.}---%
Eq.~\eqref{eq:perfect} is also known as the Knill-Laflamme (KL) condition \cite{Knill:1997kx}, and it gives the necessary and sufficient condition for quantum error detection. The standard approach for constructing error suppression codes is to start with  codes which satisfy the KL condition perfectly and design a Hamiltonian which has the code as its ground subspace \cite{PhysRevA.74.052322,Young:13}. 
However, it is well known in quantum error correction theory that relaxing the KL condition can give rise to a variety of optimized and more robust codes \cite{FletcherSW:06,Kosut:2008vl,Beny:2010zp,Ng:2010ud}. In the same spirit, we consider a relaxed error suppression condition.  Intuitively, if the KL condition is only slightly violated then the adverse effect of the noise perturbation should be small. 

To quantify this we now argue that a useful quantity which gives a simple characterization of the effect of the perturbation $V_{\lambda}$ on the code subspace is (dropping the subscript $\lambda$ for notational simplicity where possible)
\beq
\Delta E_V \equiv \max_{\ket{\psi},\ket{\phi}\in\mC} 
\left| \bra{\psi} V \ket{\psi}-\bra{\phi}  V \ket{\phi} \right|. 
\label{eq:IDS}
\eeq
This quantity determines how much the degeneracy of the code subspace $\mC$ is split due to the perturbation $V$, so we refer to $\Delta E_V$ as the ``induced degeneracy splitting" (IDS).
Remarkably, it turns out that
\beq
\min_{\alpha} \|\PC  V \PC -\alpha \PC \|= \frac{1}{2}\Delta E_V \leq \|V\|.
\label{bound3}
\eeq
In other words, $\Delta E_V$ quantifies the deviation from the KL condition. The KL condition is a special case where the IDS vanishes for all perturbations  that the code can detect. The upper bound 
follows immediately from Eq.~\eqref{eq:IDS} and the definition of the operator norm. It means that 
in the totally unprotected case where $H_{0}= 0$, so that $\mC$ is the entire Hilbert space, the IDS due to the perturbation $V$ cannot be larger than $2\|V\|$. Note that for topological codes $\Delta E_V=0$ for all (local) errors supported on regions whose size is comparable to the system size \cite{Kitaev:97,Dennis:02,Zanardi:02,Bombin:2014kx}.

We now give two complementary characterizations of the role of the IDS. We assume that $\Eg \rightarrow \infty$ so that Eq.~\eqref{eq:Heff} holds, that the state at time $t$ is given by Eq.~\eqref{eq:rho}, and that the initial state is pure and in the code space: $\rho(0)=\ketbra{\psi}{\psi}$ with $\ket{\psi}\in\mC$. Under these assumptions we find that for any perturbation $V_\lambda$
\beq
F[\ket{\psi},\rho(t)]\ge 1-t^2 \braket{(\Delta E_{V_\lambda})^2}/8\ ,
\label{eq:F}
\eeq 
where $F[\rho(t),\ket{\psi}] \equiv \sqrt{\bra{\psi}\rho(t)\ket{\psi}}$ is the fidelity \cite{nielsen2000quantum}, and the average is with respect to the random distribution $p(\lambda)$. This characterizes the IDS as setting a lower bound on the fidelity.

Our second characterization yields an upper bound on the coherence time in terms of the IDS for a special type of randomness, i.e., randomness in the strength of a fixed perturbation. Specifically, suppose the perturbation $V_\lambda =\lambda V$ where $V$ is fixed and the perturbation strength $\lambda$ is a dimensionless, random real number with probability distribution $p(\lambda)$. We call this the random-magnitude model. 
Diagonalizing $V'=\PC V \PC$ in the basis $\{|\mu_{m}\rangle\} $ with corresponding eigenvalues $\{\mu_{m}\}$, we find directly from Eq.~\eqref{eq:rho} that in this basis the matrix elements of $\rho(t)$ are given by
\begin{equation}
\rho_{mn}(t)\equiv \bra{\mu_m}\rho(t)\ket{\mu_n}  =\tilde{p}(t[\mu_{m}-\mu_{n}]) \rho_{mn}(0) ,
\label{dephas}
\end{equation}
where $\tilde{p}(\alpha) \equiv \int_{-\infty}^\infty d\lambda p(\lambda) e^{-i\lambda \alpha}$ is the Fourier transform of $p(\lambda)$.  This means that $\rho(t)$ dephases in the eigenbasis of  $\PC V \PC $, and the same function determines the magnitude of all matrix elements of $\rho(t)$. In other words, Eq.~\eqref{dephas} is a scaling relation, with $\mu_{m}-\mu_{n}$ being the scale factor. This implies that $\Delta E_{V}$ sets the decoherence timescale. To see this explicitly, define $t_\eps^{mn}$ as the time at which the coherence drops below $1-\eps$ for the first time, i.e., $|\tilde{p}(t[\mu_{m}-\mu_{n}])|  > 1-\eps$ for $0\leq t< t_\eps^{mn}$, with equality at $t=t_\eps^{mn}$. Since the RHS depends only on $\eps$, the argument must equal a constant $c_\eps$: $t_\eps^{mn} = c_\eps/(\mu_{m}-\mu_{n})$. Since $\max_{m,n}|\mu_{m}-\mu_{n}| = \Delta E_{V}$ [Eq.~\eqref{eq:IDS}], in order for the coherence $\tilde{p}(t\Delta E_{V})$ of the corresponding fastest decaying matrix element to be above $\tilde{p}(c_\eps) $, the time $t$ must be below
\beq
\tau_{\eps} \equiv \min_{m,n} t_\eps^{mn} = c_\eps/\Delta E_{V} \ ,
\label{eq:dec}
\eeq
Thus, the IDS sets an upper bound on the coherence time $\tau_{\eps}$ in the random-magnitude model. The constant $c_\eps$ is determined entirely by the properties of the distribution $p(\lambda)$. For instance, by Taylor expanding around $\alpha=0$ we have $|\tilde{p}(\alpha)|  = 1-\frac{1}{2}\textrm{var}(\lambda)\alpha^2+O(\alpha^3)$, 
and to lowest order in $\alpha$ we have $c_\eps = \sqrt{2\eps/\textrm{var}(\lambda)}$. Finally, also note that for this model $\braket{\Delta E^2_{V_\lambda}} = \braket{\lambda^2} \Delta E_V^2 $ in Eq.~\eqref{eq:F}. 

\textit{The two-site case}.---%
To make further progress in the proof of Theorem~\ref{Thm:main} let us now focus on the case of a system comprising only two sites $A$ and $B$, each supporting a qudit. The error suppression that can be obtained in this case is limited:
\begin{mylemma}
\label{lem:bound-for-2qudits}
Let $P$ be a projector in $\mathcal{H}_A\otimes  \mathcal{H}_B$ with rank larger than one. Then there exists a single-site operator $X$ such that for any complex number $\alpha$
 \begin{equation}
  \| P X P-\alpha P\| \ge \frac{\|X\|}{6}\ .
\label{eq:bound-for-2qudits}
\end{equation}
\end{mylemma}
By choosing $V=X$, it follows from Eq.~\eqref{bound3} that $\Delta E_X $, the IDS for $X$, is at least $\| X \|/3$. 

Lemma~\ref{lem:bound-for-2qudits} relies on a deep quantum phenomenon, known as the no-hiding theorem  \cite{Braunstein:2007th}, which essentially states that, unlike classical information, quantum information cannot be perfectly hidden in the correlations of two systems \cite{Cleve:99a,comment:no-hiding}. 
Here we present a stronger version of the no-hiding theorem, which is of independent interest, and which we use in our proof of Lemma~\ref{lem:bound-for-2qudits}:
 \begin{mylemma}[No-hiding]
 \label{No-hiding}
 Let $\mathcal{H}_{2}$ be an arbitrary two-dimensional subspace of $\mathcal{H}_A\otimes  \mathcal{H}_B$. Then, there exists a pair of orthonormal states $|\psi\rangle$ and $|\phi\rangle$ in  $\mathcal{H}_{2}$ for which
\end{mylemma}\color{black}
 \begin{align}
\|\Tr_{A}\left(\ketbra{\psi}{\psi}\right)&-\Tr_{A}\left(\ketbra{\phi}{\phi}\right) \|_{1}\notag \\  
 &+\|\Tr_{B}\left(\ketbra{\psi}{\psi}\right)-\Tr_{B}\left(\ketbra{\phi}{\phi}\right) \|_{1} \ge \frac{2}{3}\ .
 \label{eq:no-hiding}
 \end{align}
By Helstrom's theorem \cite{Helstrom} the trace-norm distance $\|\rho_1-\rho_2\|_1$ determines the distinguishability of the pair of states $\rho_1$ and $\rho_2$. 
Thus the more $\ket{\psi}$ and $\ket{\phi}$ become indistinguishable in $\mathcal{H}_A$, the more distinguishable they become in $\mathcal{H}_B$, and \textit{v.v.} Lemma~\ref{No-hiding} tightens the original no-hiding theorem \cite{Braunstein:2007th}, where inequality \eqref{eq:no-hiding} is stated as $>0$ \cite{comment:no-hiding2}.

Let us now sketch the proof of Lemma~\ref{lem:bound-for-2qudits}. Let $P$ be a projector onto a subspace of $\mathcal{H}_A\otimes  \mathcal{H}_B$ with dimension larger than one. From Lemma~\ref{No-hiding} we know that there exists a pair of orthonormal states in this subspace which can be distinguished by performing local measurements on either $A$ or $B$. In other words, there exists a projection operator $X$ which acts either on $A$ or $B$ such that its expectation value is different for these two orthonormal states. 
This means that $X$ does not act as the identity operator on this subspace, i.e., $P X P\not\propto P$. Indeed, using the bound in Lemma~\ref{No-hiding} there exists a projector $X$ which acts either on $A$ or $B$ for which $P X P$ is not close to $\alpha P$ for any $\alpha$; otherwise the reduced states in inequality \eqref{eq:no-hiding} would be close in the trace norm and the lower bound could not hold. This leads to Lemma~\ref{lem:bound-for-2qudits}.

\textit{Two-body commuting Hamiltonians}.---%
We now consider an arbitrary number of sites but further restrict our attention to the case of most direct relevance to real physical systems, where all the interactions are $2$-local. As we shall show, this setting has no more error suppression power than a single pair of sites. Essentially, the reason is that the ground subspaces of two-body commuting Hamiltonians have a short-range (two-body) entanglement structure that is too simple to allow for the detection of single-site errors, or even just their suppression. Technically, the decomposition of the algebra generated by two-body commuting Hamiltonians into an irreducible matrix algebra \cite{Landsman:98a} implies that $\PC$ is a direct sum of tensor products of the projectors onto the ground subspace of a non-interacting set of two-body systems. As a result we can use Lemma~\ref{lem:bound-for-2qudits} to show that:
\begin{mylemma}
\label{lemma:main}
Let $\PC $ be the projector onto the degenerate ground subspace ${\mC}$ of a Hamiltonian $H_0$ which is a sum of two-body commuting terms. Then there exists a single-site operator $X$ such that for any complex number $\alpha$
 \begin{equation}
   \|\PC  X \PC -\alpha \PC \| \ge \frac{ \|X\|}{6} \ ,
 \label{eq:1/3}
 \end{equation} which, by virtue of Eq.~\eqref{bound3}, implies that $\Delta E_X$, the IDS for $X$, is lower bounded by $\|X\|/3$. 
\end{mylemma}
Thus, any two-qudit code will always violate the KL condition in proportion to the magnitude of the perturbation. Moreover, the error suppression that can be obtained with an arbitrary number of qudits interacting via two-body commuting Hamiltonians is limited in just the same way as in the two qudits case [Eq.~\eqref{eq:bound-for-2qudits}]. 
Note that it does not help that the dimensions $d_i$ of the qudits are unrestricted. 

Now recall that in the totally unprotected case where $H_{0}= 0$, the IDS for $X$ is at most $2\| X \|$ [Eq.~\eqref{bound3}]. This only differs from the commuting two-body Hamiltonian case by a constant factor. Consequently we have established the first part of Theorem~\ref{Thm:main}, i.e., that 
for any code defined as the degenerate ground subspace of a Hamiltonian that is a sum of two-body commuting terms, there exists a single-site perturbation which induces a degeneracy splitting that is proportional to the magnitude of the perturbation, exactly as in the unprotected case.

The second part of Theorem~\ref{Thm:main} now follows from Eq.~\eqref{eq:dec}: when $H_0$ is a sum of commuting two-body terms the coherence time $\tau_\eps$ is upper-bounded by a quantity that is proportional to $\| X \|^{-1}$, just as in the totally unprotected case, when $H_0=0$. This result holds for the random-magnitude  model considered in the derivation of Eq.~\eqref{eq:dec}. However, in practice one typically has even less information about the perturbation than assumed in this model. For instance, we typically know neither the magnitude nor the exact direction of local fields that comprise the perturbation. Including additional uncertainty, such as about the direction of local fields, will only result in even tighter bounds on the coherence time. 

\textit{Discussion}.---%
It is, of course, much easier to experimentally construct and control $2$-local Hamiltonians than $k$-local Hamiltonians with $k>2$. One might thus hope that such Hamiltonians can be used to encode quantum information in a protected ground subspace. Unfortunately, we have shown that Hamiltonians consisting of two-body commuting terms (which are automatically gapped) have essentially no quantum error suppression power, even in the limit of an infinite energy gap: there always exists a single-site perturbation that can split the degeneracy of the ground subspace of such Hamiltonians, and the resulting splitting does not depend on the dimension or number of qudits, so that increasing either does not help. In other words, there always exist single-site errors causing a degeneracy splitting of the same order as the case where there is no protection whatsoever. This is a consequence of the  ground subspace of two-body commuting Hamiltonians supporting only short-range (two-body) entanglement, so that they are no more powerful than codes using only two qudits. But such codes cannot suppress arbitrary single-site errors even if the dimension of the qudits is arbitrary, as follows from the approximate version of the no-hiding theorem we proved here. 

Our results have  implications for the prospects for fault tolerant adiabatic quantum computing \cite{PhysRevA.74.052322,PhysRevLett.100.160506,Young:13,Ganti:2014db,Mizel:2014dp}, as it now evident that a passive approach that relies entirely on error suppression via a large gap generated by two-body commuting Hamiltonians will not suffice. Our no-go theorem leaves open the possibility that $2$-local Hamiltonians comprising non-commuting terms may be useful for quantum error suppression. This therefore appears as a fruitful future direction for research, if the quest for quantum error suppression using two-body interactions is to be realized, without resorting to the additional standard tools of quantum fault tolerance, such as inclusion of feedback.

\begin{acknowledgments}
This work was supported under ARO MURI Grant No. W911NF-11-1-0268 and ARO grant number W911NF-12-1-0523. The authors thank Paolo Zanardi, Fernando Pastawski and Siddharth Muthukrishnan for useful comments, and the IQIM at Caltech, where part of this work was completed, for its hospitality.  
\end{acknowledgments}


\bibliography{refs}

\begin{thebibliography}{48}
\expandafter\ifx\csname natexlab\endcsname\relax\def\natexlab#1{#1}\fi
\expandafter\ifx\csname bibnamefont\endcsname\relax
  \def\bibnamefont#1{#1}\fi
\expandafter\ifx\csname bibfnamefont\endcsname\relax
  \def\bibfnamefont#1{#1}\fi
\expandafter\ifx\csname citenamefont\endcsname\relax
  \def\citenamefont#1{#1}\fi
\expandafter\ifx\csname url\endcsname\relax
  \def\url#1{\texttt{#1}}\fi
\expandafter\ifx\csname urlprefix\endcsname\relax\def\urlprefix{URL }\fi
\providecommand{\bibinfo}[2]{#2}
\providecommand{\eprint}[2][]{\url{#2}}

\bibitem[{\citenamefont{Kitaev}(2003)}]{Kitaev:97}
\bibinfo{author}{\bibfnamefont{A.}~\bibnamefont{Kitaev}},
  \bibinfo{journal}{Ann. of Phys.} \textbf{\bibinfo{volume}{303}},
  \bibinfo{pages}{2} (\bibinfo{year}{2003}).

\bibitem[{\citenamefont{Zanardi and Rasetti}(1999)}]{HQC}
\bibinfo{author}{\bibfnamefont{P.}~\bibnamefont{Zanardi}} \bibnamefont{and}
  \bibinfo{author}{\bibfnamefont{M.}~\bibnamefont{Rasetti}},
  \bibinfo{journal}{Physics Letters A} \textbf{\bibinfo{volume}{264}},
  \bibinfo{pages}{94} (\bibinfo{year}{1999}),
  \urlprefix\url{http://www.sciencedirect.com/science/article/pii/S03759601990%
08038}.

\bibitem[{\citenamefont{Jordan et~al.}(2006)\citenamefont{Jordan, Farhi, and
  Shor}}]{PhysRevA.74.052322}
\bibinfo{author}{\bibfnamefont{S.~P.} \bibnamefont{Jordan}},
  \bibinfo{author}{\bibfnamefont{E.}~\bibnamefont{Farhi}}, \bibnamefont{and}
  \bibinfo{author}{\bibfnamefont{P.~W.} \bibnamefont{Shor}},
  \bibinfo{journal}{Phys. Rev. A} \textbf{\bibinfo{volume}{74}},
  \bibinfo{pages}{052322} (\bibinfo{year}{2006}),
  \urlprefix\url{http://link.aps.org/doi/10.1103/PhysRevA.74.052322}.

\bibitem[{\citenamefont{Bravyi and Vyalyi}(2005)}]{Bravyi03}
\bibinfo{author}{\bibfnamefont{S.}~\bibnamefont{Bravyi}} \bibnamefont{and}
  \bibinfo{author}{\bibfnamefont{M.}~\bibnamefont{Vyalyi}},
  \bibinfo{journal}{Quantum Inf. and Comp.} \textbf{\bibinfo{volume}{5}},
  \bibinfo{pages}{187} (\bibinfo{year}{2005}).

\bibitem[{\citenamefont{Bravyi et~al.}(2010)\citenamefont{Bravyi, Poulin, and
  Terhal}}]{BPT_Com}
\bibinfo{author}{\bibfnamefont{S.}~\bibnamefont{Bravyi}},
  \bibinfo{author}{\bibfnamefont{D.}~\bibnamefont{Poulin}}, \bibnamefont{and}
  \bibinfo{author}{\bibfnamefont{B.}~\bibnamefont{Terhal}},
  \bibinfo{journal}{Physical Review Letters} \textbf{\bibinfo{volume}{104}},
  \bibinfo{pages}{050503} (\bibinfo{year}{2010}),
  \urlprefix\url{http://link.aps.org/doi/10.1103/PhysRevLett.104.050503}.

\bibitem[{\citenamefont{Aharonov and Eldar}(2011)}]{Ahar11}
\bibinfo{author}{\bibfnamefont{D.}~\bibnamefont{Aharonov}} \bibnamefont{and}
  \bibinfo{author}{\bibfnamefont{L.}~\bibnamefont{Eldar}},
  \bibinfo{journal}{Foundations of Computer Science (FOCS), IEEE 52nd Annual
  Symposium} pp. \bibinfo{pages}{334--343} (\bibinfo{year}{2011}).

\bibitem[{\citenamefont{Hastings}(2011)}]{Hastings_Top}
\bibinfo{author}{\bibfnamefont{M.~B.} \bibnamefont{Hastings}},
  \bibinfo{journal}{Physical Review Letters} \textbf{\bibinfo{volume}{107}},
  \bibinfo{pages}{210501} (\bibinfo{year}{2011}),
  \urlprefix\url{http://link.aps.org/doi/10.1103/PhysRevLett.107.210501}.

\bibitem[{\citenamefont{Haah and Preskill}(2012)}]{Haah_Preskill}
\bibinfo{author}{\bibfnamefont{J.}~\bibnamefont{Haah}} \bibnamefont{and}
  \bibinfo{author}{\bibfnamefont{J.}~\bibnamefont{Preskill}},
  \bibinfo{journal}{Physical Review A} \textbf{\bibinfo{volume}{86}},
  \bibinfo{pages}{032308} (\bibinfo{year}{2012}),
  \urlprefix\url{http://link.aps.org/doi/10.1103/PhysRevA.86.032308}.

\bibitem[{\citenamefont{Gottesman}(1996)}]{Gottesman:1996fk}
\bibinfo{author}{\bibfnamefont{D.}~\bibnamefont{Gottesman}},
  \bibinfo{journal}{{Phys. Rev. A}} \textbf{\bibinfo{volume}{54}},
  \bibinfo{pages}{1862} (\bibinfo{year}{1996}).

\bibitem[{\citenamefont{Dennis et~al.}(2002)\citenamefont{Dennis, Kitaev,
  Landahl, and Preskill}}]{Dennis:02}
\bibinfo{author}{\bibfnamefont{E.}~\bibnamefont{Dennis}},
  \bibinfo{author}{\bibfnamefont{A.}~\bibnamefont{Kitaev}},
  \bibinfo{author}{\bibfnamefont{A.}~\bibnamefont{Landahl}}, \bibnamefont{and}
  \bibinfo{author}{\bibfnamefont{J.}~\bibnamefont{Preskill}},
  \bibinfo{journal}{Journal of Mathematical Physics}
  \textbf{\bibinfo{volume}{43}}, \bibinfo{pages}{4452} (\bibinfo{year}{2002}),
  \urlprefix\url{http://scitation.aip.org/content/aip/journal/jmp/43/9/10.1063%
/1.1499754}.

\bibitem[{\citenamefont{Bomb{\'\i}n}(2014)}]{Bombin:2014kx}
\bibinfo{author}{\bibfnamefont{H.}~\bibnamefont{Bomb{\'\i}n}},
  \bibinfo{journal}{Commun. Math. Phys.} \textbf{\bibinfo{volume}{327}},
  \bibinfo{pages}{387} (\bibinfo{year}{2014}),
  \urlprefix\url{http://dx.doi.org/10.1007/s00220-014-1893-4}.

\bibitem[{\citenamefont{Bacon}(2006)}]{Bacon:05}
\bibinfo{author}{\bibfnamefont{D.}~\bibnamefont{Bacon}},
  \bibinfo{journal}{Physical Review A} \textbf{\bibinfo{volume}{73}},
  \bibinfo{pages}{012340} (\bibinfo{year}{2006}),
  \urlprefix\url{http://link.aps.org/doi/10.1103/PhysRevA.73.012340}.

\bibitem[{\citenamefont{Chesi et~al.}(2010)\citenamefont{Chesi, Loss, Bravyi,
  and Terhal}}]{Chesi:2010rc}
\bibinfo{author}{\bibfnamefont{S.}~\bibnamefont{Chesi}},
  \bibinfo{author}{\bibfnamefont{D.}~\bibnamefont{Loss}},
  \bibinfo{author}{\bibfnamefont{S.}~\bibnamefont{Bravyi}}, \bibnamefont{and}
  \bibinfo{author}{\bibfnamefont{B.~M.} \bibnamefont{Terhal}},
  \bibinfo{journal}{New Journal of Physics} \textbf{\bibinfo{volume}{12}},
  \bibinfo{pages}{025013} (\bibinfo{year}{2010}),
  \urlprefix\url{http://stacks.iop.org/1367-2630/12/i=2/a=025013}.

\bibitem[{\citenamefont{Haah}(2011)}]{Haah:2011fv}
\bibinfo{author}{\bibfnamefont{J.}~\bibnamefont{Haah}},
  \bibinfo{journal}{Physical Review A} \textbf{\bibinfo{volume}{83}},
  \bibinfo{pages}{042330} (\bibinfo{year}{2011}),
  \urlprefix\url{http://link.aps.org/doi/10.1103/PhysRevA.83.042330}.

\bibitem[{\citenamefont{Hastings et~al.}(2014)\citenamefont{Hastings, Watson,
  and Melko}}]{Hastings:2014uq}
\bibinfo{author}{\bibfnamefont{M.~B.} \bibnamefont{Hastings}},
  \bibinfo{author}{\bibfnamefont{G.~H.} \bibnamefont{Watson}},
  \bibnamefont{and} \bibinfo{author}{\bibfnamefont{R.~G.} \bibnamefont{Melko}},
  \bibinfo{journal}{Physical Review Letters} \textbf{\bibinfo{volume}{112}},
  \bibinfo{pages}{070501} (\bibinfo{year}{2014}),
  \urlprefix\url{http://link.aps.org/doi/10.1103/PhysRevLett.112.070501}.

\bibitem[{\citenamefont{Lidar}(2008)}]{PhysRevLett.100.160506}
\bibinfo{author}{\bibfnamefont{D.~A.} \bibnamefont{Lidar}},
  \bibinfo{journal}{{Phys.~Rev.~Lett.}} \textbf{\bibinfo{volume}{100}},
  \bibinfo{pages}{160506} (\bibinfo{year}{2008}),
  \urlprefix\url{http://link.aps.org/doi/10.1103/PhysRevLett.100.160506}.

\bibitem[{\citenamefont{Young et~al.}(2013)\citenamefont{Young, Sarovar, and
  Blume-Kohout}}]{Young:13}
\bibinfo{author}{\bibfnamefont{K.~C.} \bibnamefont{Young}},
  \bibinfo{author}{\bibfnamefont{M.}~\bibnamefont{Sarovar}}, \bibnamefont{and}
  \bibinfo{author}{\bibfnamefont{R.}~\bibnamefont{Blume-Kohout}},
  \bibinfo{journal}{Physical Review X} \textbf{\bibinfo{volume}{3}},
  \bibinfo{pages}{041013} (\bibinfo{year}{2013}),
  \urlprefix\url{http://link.aps.org/doi/10.1103/PhysRevX.3.041013}.

\bibitem[{\citenamefont{Ganti et~al.}(2014)\citenamefont{Ganti, Onunkwo, and
  Young}}]{Ganti:2014db}
\bibinfo{author}{\bibfnamefont{A.}~\bibnamefont{Ganti}},
  \bibinfo{author}{\bibfnamefont{U.}~\bibnamefont{Onunkwo}}, \bibnamefont{and}
  \bibinfo{author}{\bibfnamefont{K.}~\bibnamefont{Young}},
  \bibinfo{journal}{Physical Review A} \textbf{\bibinfo{volume}{89}},
  \bibinfo{pages}{042313} (\bibinfo{year}{2014}),
  \urlprefix\url{http://link.aps.org/doi/10.1103/PhysRevA.89.042313}.

\bibitem[{\citenamefont{Mizel}(2014)}]{Mizel:2014dp}
\bibinfo{author}{\bibfnamefont{A.}~\bibnamefont{Mizel}},
  \bibinfo{journal}{arXiv:1403.7694}  (\bibinfo{year}{2014}),
  \urlprefix\url{http://arXiv.org/abs/1403.7694}.

\bibitem[{\citenamefont{Gottesman et~al.}(2001)\citenamefont{Gottesman, Kitaev,
  and Preskill}}]{Got2}
\bibinfo{author}{\bibfnamefont{D.}~\bibnamefont{Gottesman}},
  \bibinfo{author}{\bibfnamefont{A.}~\bibnamefont{Kitaev}}, \bibnamefont{and}
  \bibinfo{author}{\bibfnamefont{J.}~\bibnamefont{Preskill}},
  \bibinfo{journal}{Physical Review A} \textbf{\bibinfo{volume}{64}},
  \bibinfo{pages}{012310} (\bibinfo{year}{2001}),
  \urlprefix\url{http://link.aps.org/doi/10.1103/PhysRevA.64.012310}.

\bibitem[{\citenamefont{Bravyi and Terhal}(2009)}]{Bravyi:2009rz}
\bibinfo{author}{\bibfnamefont{S.}~\bibnamefont{Bravyi}} \bibnamefont{and}
  \bibinfo{author}{\bibfnamefont{B.}~\bibnamefont{Terhal}},
  \bibinfo{journal}{New Journal of Physics} \textbf{\bibinfo{volume}{11}},
  \bibinfo{pages}{043029} (\bibinfo{year}{2009}),
  \urlprefix\url{http://stacks.iop.org/1367-2630/11/i=4/a=043029}.

\bibitem[{\citenamefont{Pastawski et~al.}(2010)\citenamefont{Pastawski, Kay,
  Schuch, and Cirac}}]{Pastawski:2009fk}
\bibinfo{author}{\bibfnamefont{F.}~\bibnamefont{Pastawski}},
  \bibinfo{author}{\bibfnamefont{A.}~\bibnamefont{Kay}},
  \bibinfo{author}{\bibfnamefont{N.}~\bibnamefont{Schuch}}, \bibnamefont{and}
  \bibinfo{author}{\bibfnamefont{I.}~\bibnamefont{Cirac}},
  \bibinfo{journal}{Quantum Inf. and Comp.} \textbf{\bibinfo{volume}{10}},
  \bibinfo{pages}{580} (\bibinfo{year}{2010}),
  \urlprefix\url{http://arXiv.org/abs/0911.3843}.

\bibitem[{\citenamefont{Yoshida}(2011)}]{Yoshida:2011fk}
\bibinfo{author}{\bibfnamefont{B.}~\bibnamefont{Yoshida}},
  \bibinfo{journal}{Annals of Physics} \textbf{\bibinfo{volume}{326}},
  \bibinfo{pages}{2566} (\bibinfo{year}{2011}),
  \urlprefix\url{http://www.sciencedirect.com/science/article/pii/S00034916110%
01023}.

\bibitem[{\citenamefont{Kribs et~al.}(2005)\citenamefont{Kribs, Laflamme, and
  Poulin}}]{Kribs:2005:180501}
\bibinfo{author}{\bibfnamefont{D.}~\bibnamefont{Kribs}},
  \bibinfo{author}{\bibfnamefont{R.}~\bibnamefont{Laflamme}}, \bibnamefont{and}
  \bibinfo{author}{\bibfnamefont{D.}~\bibnamefont{Poulin}},
  \bibinfo{journal}{Phys. Rev. Lett.} \textbf{\bibinfo{volume}{94}},
  \bibinfo{pages}{180501} (\bibinfo{year}{2005}),
  \urlprefix\url{http://link.aps.org/doi/10.1103/PhysRevLett.94.180501}.

\bibitem[{\citenamefont{Lidar and Brun}(2013)}]{Lidar-Brun:book}
\bibinfo{editor}{\bibfnamefont{D.}~\bibnamefont{Lidar}} \bibnamefont{and}
  \bibinfo{editor}{\bibfnamefont{T.}~\bibnamefont{Brun}}, eds.,
  \emph{\bibinfo{title}{Quantum Error Correction}}
  (\bibinfo{publisher}{Cambridge University Press},
  \bibinfo{address}{{Cambride, UK}}, \bibinfo{year}{2013}),
  \urlprefix\url{http://www.cambridge.org/9780521897877}.

\bibitem[{\citenamefont{Cleve et~al.}(1999)\citenamefont{Cleve, Gottesman, and
  Lo}}]{Cleve:99a}
\bibinfo{author}{\bibfnamefont{R.}~\bibnamefont{Cleve}},
  \bibinfo{author}{\bibfnamefont{D.}~\bibnamefont{Gottesman}},
  \bibnamefont{and} \bibinfo{author}{\bibfnamefont{H.-K.} \bibnamefont{Lo}},
  \bibinfo{journal}{Physical Review Letters} \textbf{\bibinfo{volume}{83}},
  \bibinfo{pages}{648} (\bibinfo{year}{1999}),
  \urlprefix\url{http://link.aps.org/doi/10.1103/PhysRevLett.83.648}.

\bibitem[{\citenamefont{Braunstein and Pati}(2007)}]{Braunstein:2007th}
\bibinfo{author}{\bibfnamefont{S.~L.} \bibnamefont{Braunstein}}
  \bibnamefont{and} \bibinfo{author}{\bibfnamefont{A.~K.} \bibnamefont{Pati}},
  \bibinfo{journal}{Physical Review Letters} \textbf{\bibinfo{volume}{98}},
  \bibinfo{pages}{080502} (\bibinfo{year}{2007}),
  \urlprefix\url{http://link.aps.org/doi/10.1103/PhysRevLett.98.080502}.

\bibitem[{\citenamefont{Kretschmann et~al.}(2008)\citenamefont{Kretschmann,
  Kribs, and Spekkens}}]{Kretschmann:2008zp}
\bibinfo{author}{\bibfnamefont{D.}~\bibnamefont{Kretschmann}},
  \bibinfo{author}{\bibfnamefont{D.~W.} \bibnamefont{Kribs}}, \bibnamefont{and}
  \bibinfo{author}{\bibfnamefont{R.~W.} \bibnamefont{Spekkens}},
  \bibinfo{journal}{Physical Review A} \textbf{\bibinfo{volume}{78}},
  \bibinfo{pages}{032330} (\bibinfo{year}{2008}),
  \urlprefix\url{http://link.aps.org/doi/10.1103/PhysRevA.78.032330}.

\bibitem[{\citenamefont{Audenaert and Scheel}(2008)}]{Audenaert:08}
\bibinfo{author}{\bibfnamefont{K.~M.~R.} \bibnamefont{Audenaert}}
  \bibnamefont{and} \bibinfo{author}{\bibfnamefont{S.}~\bibnamefont{Scheel}},
  \bibinfo{journal}{New Journal of Physics} \textbf{\bibinfo{volume}{10}},
  \bibinfo{pages}{023011} (\bibinfo{year}{2008}),
  \urlprefix\url{http://stacks.iop.org/1367-2630/10/i=2/a=023011}.

\bibitem[{\citenamefont{Marvian and Lidar}()}]{ML:14}
\bibinfo{author}{\bibfnamefont{I.}~\bibnamefont{Marvian}} \bibnamefont{and}
  \bibinfo{author}{\bibfnamefont{D.}~\bibnamefont{Lidar}}, \bibinfo{note}{to be
  published}.

\bibitem[{\citenamefont{Zanardi and Venuti}(2014)}]{Zanardi:2014fr}
\bibinfo{author}{\bibfnamefont{P.}~\bibnamefont{Zanardi}} \bibnamefont{and}
  \bibinfo{author}{\bibfnamefont{L.~C.} \bibnamefont{Venuti}},
  \bibinfo{journal}{arXiv:1404.4673}  (\bibinfo{year}{2014}),
  \urlprefix\url{http://arXiv.org/abs/1404.4673}.

\bibitem[{\citenamefont{Bookatz et~al.}(2014)\citenamefont{Bookatz, Farhi, and
  Zhou}}]{Bookatz:2014uq}
\bibinfo{author}{\bibfnamefont{A.~D.} \bibnamefont{Bookatz}},
  \bibinfo{author}{\bibfnamefont{E.}~\bibnamefont{Farhi}}, \bibnamefont{and}
  \bibinfo{author}{\bibfnamefont{L.}~\bibnamefont{Zhou}},
  \bibinfo{journal}{arXiv:1407.1485}  (\bibinfo{year}{2014}),
  \urlprefix\url{http://arXiv.org/abs/1407.1485}.

\bibitem[{\citenamefont{Suzuki and Okamoto}(1983)}]{Suzuki:1983cs}
\bibinfo{author}{\bibfnamefont{K.}~\bibnamefont{Suzuki}} \bibnamefont{and}
  \bibinfo{author}{\bibfnamefont{R.}~\bibnamefont{Okamoto}},
  \bibinfo{journal}{Progress of Theoretical Physics}
  \textbf{\bibinfo{volume}{70}}, \bibinfo{pages}{439} (\bibinfo{year}{1983}),
  \urlprefix\url{http://ptp.oxfordjournals.org/content/70/2/439.abstract}.

\bibitem[{\citenamefont{Knill and Laflamme}(1997)}]{Knill:1997kx}
\bibinfo{author}{\bibfnamefont{E.}~\bibnamefont{Knill}} \bibnamefont{and}
  \bibinfo{author}{\bibfnamefont{R.}~\bibnamefont{Laflamme}},
  \bibinfo{journal}{Physical Review A} \textbf{\bibinfo{volume}{55}},
  \bibinfo{pages}{900} (\bibinfo{year}{1997}),
  \urlprefix\url{http://link.aps.org/doi/10.1103/PhysRevA.55.900}.

\bibitem[{\citenamefont{Fletcher et~al.}(2007)\citenamefont{Fletcher, Shor, and
  Win}}]{FletcherSW:06}
\bibinfo{author}{\bibfnamefont{A.~S.} \bibnamefont{Fletcher}},
  \bibinfo{author}{\bibfnamefont{P.~W.} \bibnamefont{Shor}}, \bibnamefont{and}
  \bibinfo{author}{\bibfnamefont{M.~Z.} \bibnamefont{Win}},
  \bibinfo{journal}{Physical Review A} \textbf{\bibinfo{volume}{75}},
  \bibinfo{pages}{012338} (\bibinfo{year}{2007}),
  \urlprefix\url{http://link.aps.org/doi/10.1103/PhysRevA.75.012338}.

\bibitem[{\citenamefont{Kosut et~al.}(2008)\citenamefont{Kosut, Shabani, and
  Lidar}}]{Kosut:2008vl}
\bibinfo{author}{\bibfnamefont{R.~L.} \bibnamefont{Kosut}},
  \bibinfo{author}{\bibfnamefont{A.}~\bibnamefont{Shabani}}, \bibnamefont{and}
  \bibinfo{author}{\bibfnamefont{D.~A.} \bibnamefont{Lidar}},
  \bibinfo{journal}{Physical Review Letters} \textbf{\bibinfo{volume}{100}},
  \bibinfo{pages}{020502} (\bibinfo{year}{2008}),
  \urlprefix\url{http://link.aps.org/doi/10.1103/PhysRevLett.100.020502}.

\bibitem[{\citenamefont{B{\'e}ny and Oreshkov}(2010)}]{Beny:2010zp}
\bibinfo{author}{\bibfnamefont{C.}~\bibnamefont{B{\'e}ny}} \bibnamefont{and}
  \bibinfo{author}{\bibfnamefont{O.}~\bibnamefont{Oreshkov}},
  \bibinfo{journal}{Physical Review Letters} \textbf{\bibinfo{volume}{104}},
  \bibinfo{pages}{120501} (\bibinfo{year}{2010}),
  \urlprefix\url{http://link.aps.org/doi/10.1103/PhysRevLett.104.120501}.

\bibitem[{\citenamefont{Ng and Mandayam}(2010)}]{Ng:2010ud}
\bibinfo{author}{\bibfnamefont{H.~K.} \bibnamefont{Ng}} \bibnamefont{and}
  \bibinfo{author}{\bibfnamefont{P.}~\bibnamefont{Mandayam}},
  \bibinfo{journal}{Physical Review A} \textbf{\bibinfo{volume}{81}},
  \bibinfo{pages}{062342} (\bibinfo{year}{2010}),
  \urlprefix\url{http://link.aps.org/doi/10.1103/PhysRevA.81.062342}.

\bibitem[{\citenamefont{Zanardi and Lloyd}(2003)}]{Zanardi:02}
\bibinfo{author}{\bibfnamefont{P.}~\bibnamefont{Zanardi}} \bibnamefont{and}
  \bibinfo{author}{\bibfnamefont{S.}~\bibnamefont{Lloyd}},
  \bibinfo{journal}{Phys. Rev. Lett.} \textbf{\bibinfo{volume}{90}},
  \bibinfo{pages}{067902} (\bibinfo{year}{2003}),
  \urlprefix\url{http://link.aps.org/doi/10.1103/PhysRevLett.90.067902}.

\bibitem[{\citenamefont{Nielsen and Chuang}(2000)}]{nielsen2000quantum}
\bibinfo{author}{\bibfnamefont{M.}~\bibnamefont{Nielsen}} \bibnamefont{and}
  \bibinfo{author}{\bibfnamefont{I.}~\bibnamefont{Chuang}},
  \emph{\bibinfo{title}{Quantum Computation and Quantum Information}},
  Cambridge Series on Information and the Natural Sciences
  (\bibinfo{publisher}{Cambridge University Press}, \bibinfo{year}{2000}), ISBN
  \bibinfo{isbn}{9780521635035}.

\bibitem[{com({\natexlab{a}})}]{comment:no-hiding}
\bibinfo{howpublished}{{The no-hiding theorem seems intuitive from the point of
  view of the theory of quantum error correction. Suppose the reduced state of
  subsystem $B$ [$\Tr_A(\ketbra{\psi}{\psi})$] is independent of the encoded
  state $|\psi\rangle$ for all states in a code subspace of
  $\mathcal{H}_A\otimes \mathcal{H}_B$. This implies that the error detection
  condition is satisfied for all errors on $B$ and so the encoded state can be
  recovered by acting just on $A$. This means that the reduced state of $A$
  [$\Tr_B(\ketbra{\psi}{\psi})$] cannot be independent of the state
  $|\psi\rangle$ in the code subspace, which implies the no-hiding theorem.}}

\bibitem[{\citenamefont{{C. W. Helstrom}}(1976)}]{Helstrom}
\bibinfo{author}{\bibnamefont{{C. W. Helstrom}}},
  \emph{\bibinfo{title}{{Quantum Detection and Estimation Theory}}}
  (\bibinfo{publisher}{{Academic Press}}, \bibinfo{address}{{New York}},
  \bibinfo{year}{1976}).

\bibitem[{com({\natexlab{b}})}]{comment:no-hiding2}
\bibinfo{howpublished}{{We note that another approximate version of the
  no-hiding theorem was found in Ref.~\cite{Kretschmann:2008zp}: if a channel
  is close to a deletion channel (a channel whose output is independent of its
  input) in the diamond norm, then its complementary channel (the channel
  obtained by tracing out the other subsystem \cite{Devetak:2005kl}) should be
  close to the identity channel in the diamond norm.}}

\bibitem[{\citenamefont{Landsman}(1998)}]{Landsman:98a}
\bibinfo{author}{\bibfnamefont{N.~P.} \bibnamefont{Landsman}},
  \bibinfo{journal}{arXiv:math-ph/9807030}  (\bibinfo{year}{1998}),
  \urlprefix\url{http://arXiv.org/abs/math-ph/9807030}.

\bibitem[{\citenamefont{Ng et~al.}(2011)\citenamefont{Ng, Lidar, and
  Preskill}}]{NLP:09}
\bibinfo{author}{\bibfnamefont{H.-K.} \bibnamefont{Ng}},
  \bibinfo{author}{\bibfnamefont{D.~A.} \bibnamefont{Lidar}}, \bibnamefont{and}
  \bibinfo{author}{\bibfnamefont{J.~P.} \bibnamefont{Preskill}},
  \bibinfo{journal}{Phys. Rev. A}  (\bibinfo{year}{2011}).

\bibitem[{\citenamefont{{R. Bhatia}}(1997)}]{Bhatia:book}
\bibinfo{author}{\bibnamefont{{R. Bhatia}}}, \emph{\bibinfo{title}{{Matrix
  Analysis}}}, no. \bibinfo{number}{169} in \bibinfo{series}{{Graduate Texts in
  Mathematics}} (\bibinfo{publisher}{Springer-Verlag}, \bibinfo{address}{{New
  York}}, \bibinfo{year}{1997}).

\bibitem[{\citenamefont{{C.A. Fuchs and J. van de Graaf}}(1999)}]{Fuchs:99}
\bibinfo{author}{\bibnamefont{{C.A. Fuchs and J. van de Graaf}}},
  \bibinfo{journal}{{IEEE Transactions on Information Theory}}
  \textbf{\bibinfo{volume}{45}}, \bibinfo{pages}{1216} (\bibinfo{year}{1999}).

\bibitem[{\citenamefont{Zanardi}(2001)}]{Zanardi:2001b}
\bibinfo{author}{\bibfnamefont{P.}~\bibnamefont{Zanardi}},
  \bibinfo{journal}{Physical Review Letters} \textbf{\bibinfo{volume}{87}},
  \bibinfo{pages}{077901} (\bibinfo{year}{2001}),
  \urlprefix\url{http://link.aps.org/doi/10.1103/PhysRevLett.87.077901}.

\end{thebibliography}

\clearpage

\appendix

\begin{center}
\large{Appendix}
\end{center}

\section{Random unitary channels from open system dynamics}
We demonstrate that open system dynamics can give rise to the random unitary channel in Eq.~\eqref{eq:rho}.  Let $H_S$ and $H_B$, respectively, be the system and bath Hamiltonians. Let $\{|\lambda\rangle\}$ be an  orthonormal basis for the bath Hilbert space with the property that $[H_B,|\lambda\rangle\langle\lambda|]=0, \ \forall |\lambda\rangle$. (For instance, the set $\{|\lambda\rangle\}$ could be eigenstates of an observable $\Lambda=\sum_\lambda \lambda |\lambda\rangle\langle\lambda|$ which commutes with the bath Hamiltonian, i.e., $[H_{B},\Lambda]=0$.)   Assume that the system-bath interaction has the following form:
\beq
H_{I}=\sum_\lambda V_\lambda\otimes |\lambda\rangle\langle\lambda|\ .
\eeq
Then, 
\bes
\begin{align}
e^{-i t (H_S+H_B+H_{I})}&=e^{-i t (H_S+H_{I})}\otimes e^{-i t H_B}\\ &= \left(\sum_\lambda e^{-i t (H_S+V_\lambda)}\otimes |\lambda\rangle\langle\lambda|\right) e^{-i t H_B}\\ &= \sum_\lambda e^{-i t (H_S+V_\lambda)}\otimes  (e^{-i t H_B}|\lambda\rangle\langle\lambda|)\label{Bath}
\end{align} 
\ees
where the first equality follows from the fact that both $H_S$ and $H_{I}$ commute with $H_B$, and the second equality follows, e.g., from the Taylor expansion of the exponential along with the completeness and orthonormality of the set $\{|\lambda\rangle\}$.

Let $\rho(0)$ be the initial state of the system and  $\rho_{SB}(0)=\rho(0)\otimes \rho_B$ be the initial joint state of the system and bath. Assume that the initial state of the bath also satisfies $[\rho_B,|\lambda\rangle\langle\lambda|]=0, \ \forall |\lambda\rangle$, so that the bath state is stationary. This will be the case, e.g., if the bath is initially in a thermal state, i.e., $\rho_B=e^{-\beta H_B}/\Tr [e^{-\beta H_B}]$.
Then, using Eq.~\eqref{Bath}, one can easily see  that the joint state of the system and bath at time $t$ is given by
\bes
\begin{align}
\rho_{SB}(t)&=e^{-i t (H_S+H_B+H_{I})}(\rho\otimes\rho_B)e^{i t (H_S+H_B+H_{I})} \\ 
&= \sum_\lambda e^{-i t (H_S+V_\lambda)}\rho\ e^{i t (H_S+V_\lambda)}\otimes |\lambda\rangle\langle\lambda|\rho_B|\lambda\rangle\langle\lambda|\\ 
&= \sum_\lambda p(\lambda) e^{-i t (H_S+V_\lambda)}\rho e^{i t (H_S+V_\lambda)} \otimes |\lambda\rangle\langle\lambda|
\end{align} 
\ees
where $p(\lambda)\equiv \langle\lambda|\rho_B|\lambda\rangle$. So, tracing over the bath we find that  $\rho(t)$, the state of system at time $t$ is given by
\begin{equation}
\rho(t)=\sum_\lambda p(\lambda)  e^{-i t (H_S+V_\lambda)}\rho e^{i t (H_S+V_\lambda)}.
\end{equation}
In other words, the evolution of system is described by a random unitary channel in which, with probability $p(\lambda)$, the Hamiltonian $H_S+V_\lambda$ is applied to the system for time $t$.  


\section{Derivation of Eq.~\eqref{bound3}}

We shall prove that 
\bes
\begin{align}
\Delta E_V &\equiv \max_{\ket{\psi},\ket{\phi}\in\mC} 
\left| \bra{\psi} V \ket{\psi}-\bra{\phi}  V \ket{\phi} \right| \\
&=
2\min_{\alpha} \|\PC  V \PC -\alpha \PC \| .
\end{align}
\ees

Let $\lambda_{\max}$ and $\lambda_{\min}$ be respectively the maximal and minimal eigenvalues of $\PC  V \PC $ whose corresponding eigenvectors live in the subspace $\mathcal{C}$. Then clearly 
\beq
\max_{\ket{\psi},\ket{\phi}\in\mC} \left| \bra{\psi} V\ket{\psi}-\bra{\phi} V\ket{\phi} \right|=\lambda_{\max}-\lambda_{\min}.
\label{eq:V-diff}
\eeq 
On the other hand,  the maximal and minimal  eigenvalues of $\PC  V \PC -\alpha \PC $ are $\lambda_{\max}-\alpha$ and $\lambda_{\min}-\alpha$, respectively. Thus
\beq
\|\PC  V \PC -\alpha \PC \| = \max\{|\lambda_{\max}-\alpha|,|\lambda_{\min}-\alpha|\} 
\eeq
It is easy to see that $\max\{|\lambda_{\max}-\alpha|,|\lambda_{\min}-\alpha|\} $ is minimized for $\alpha = \frac{1}{2}(\lambda_{\max}+\lambda_{\min})$ and hence that 
\beq
\min_\alpha \max\{|\lambda_{\max}-\alpha |,|\lambda_{\min}-\alpha |\} = \frac{1}{2}(\lambda_{\max}-\lambda_{\min})\ .
\label{eq:minmax}
\eeq
Combining Eqs.~\eqref{eq:V-diff}-\eqref{eq:minmax} proves the claim.


\section{Two characterizations of the IDS}

\subsection{First characterization: Eq.~\eqref{eq:F}}
Suppose, as in the main text, that $H_\lambda = H_{0}+V_\lambda$ and the perturbation $V_{\lambda}$ depends on some unknown parameter $\lambda$ with probability distribution $p(\lambda)$. Then, subject to the evolution operator $U_{\lambda}(t) = \exp(-it H_\lambda)$, after time $t$ the initial state $\ket{\psi}$ evolves to    
 \begin{equation}
\rho(t)=\int d\lambda\ p(\lambda)\  \ketbra{\psi_\lambda(t)}{\psi_\lambda(t)}\ ,
\end{equation}
where $\ket{\psi_\lambda(t)} = U_{\lambda}(t)\ket{\psi}$. For any state $\ket{\psi}$ in the ground subspace of $H_0$, i.e., $\PC \ket{\psi} = \ket{\psi}$, we have 
\begin{align}
\label{eq:19}
&\|U_\lambda \ket{\psi}-e^{-i t \PC V_\lambda \PC }\ket{\psi}\|  \\
&\quad \le \|U_\lambda \PC -e^{-i t \PC V_\lambda \PC }\PC \|  \le \frac{4\|V_\lambda\|}{\Eg}(\|V_\lambda\| |t|+1)\ , \notag
\end{align}
where in the first inequality we used the definition of the operator norm and in the second inequality  we used Lemma~\ref{lemma_pert}.

Note that for any pair of Hermitian operators $H_1$ and $H_2$ (e.g., \cite{NLP:09}), 
\beq
\|e^{-iH_1 t}-e^{-iH_2 t}\|\le t \|H_1-H_2\| \ .
\label{eq:expH}
\eeq
Let $\lambda_{\min}$ and $\lambda_{\max}$ be the minimum and maximum eigenvalues, respectively, of $\PC V_\lambda \PC $ in the support of $\PC $. 
Then, using Eq.~\eqref{eq:expH},
\bes
\label{eq:21}
\begin{align}
& \| e^{-i t \PC V_\lambda \PC } - e^{-it \PC \frac{\lambda_{\min}+\lambda_{\max}}{2}\PC}\| \notag \\
\label{eq:21a}
&\qquad \le t \| \PC V_\lambda \PC - \PC \frac{\lambda_{\min}+\lambda_{\max}}{2}\PC \|\\ 
\label{eq:21b}
&\qquad = t \frac{\lambda_{\max}-\lambda_{\min}}{2} =\frac{t \Delta E_{V_\lambda}}{2}\ ,
\end{align}
\ees 
where Eq.~\eqref{eq:21b} follows from the fact that the maximum eigenvalue of $\PC(V_{\lambda}-\frac{\lambda_{\max}+\lambda_{\min}}{2}I)\PC$ is $\lambda_{\max}-\frac{\lambda_{\max}+\lambda_{\min}}{2}$, which equals the IDS by its definition [Eq.~\eqref{eq:IDS}]. Therefore, combining Eqs.~\eqref{eq:19} and \eqref{eq:21} and using the triangle inequality we find that
\bes
\begin{align}
&\|U_\lambda \ket{\psi}-e^{-it \frac{\lambda_{\min}+\lambda_{\max}}{2}}\ket{\psi}\| \notag \\
&\qquad \le \|U_\lambda \ket{\psi}-e^{-i t \PC V_\lambda \PC }\ket{\psi}\|+ \notag \\ 
&\qquad \qquad\qquad \| e^{-i t \PC V_\lambda \PC } - e^{-it \PC \frac{\lambda_{\min}+\lambda_{\max}}{2}\PC}\|\\ 
&\qquad \le  \frac{4\|V_\lambda\|}{\Eg}(\|V_\lambda\| |t|+1) +\frac{t \Delta E_{V_\lambda}}{2}\ .
\end{align}
\ees 
Using the fact that for any pair of states $\ket{\phi_1}$ and $\ket{\phi_2}$,
$ 1- \left|\bracket{\phi_1}{\phi_2}\right|\le \|\ket{\phi_1}-\ket{\phi_2}\|^2/2$, this implies 
\begin{align}
\sqrt{1-|\bra{\psi}U_\lambda(t)\ket{\psi}|}\le\frac{1}{\sqrt{2}} \frac{4\|V_\lambda\|}{\Eg}(\|V_\lambda\| |t|+1) +\frac{t \Delta E_{V_\lambda}}{2\sqrt{2}}\ .
\end{align}
This bound holds for arbitrary $\Eg$.  In the limit where $\Eg\rightarrow \infty$, we find that 
\begin{align}
\label{app_proof2}
|\bra{\psi}U_\lambda(t)\ket{\psi}| \ge  1-\frac{t^2 \Delta E^2_{V_\lambda}}{8}\ .
\end{align}
It follows that for the fidelity $F[\rho(t),\ket{\psi}]=\sqrt{\bra{\psi}\rho(t)\ket{\psi}}$:
\bes
\begin{align}
F[\rho(t),\ket{\psi}] &=\sqrt{\int d\lambda\  p(\lambda) |\bra{\psi}U_\lambda(t)\ket{\psi}|^2}  \\ 
\label{eq:25b}
&\ge \int d\lambda\  p(\lambda) |\bra{\psi}U_\lambda(t)\ket{\psi}|\\ 
\label{eq:25c}
&\ge \int d\lambda\  p(\lambda) \left[1-\frac{\left(t\Delta E_{V_\lambda}\right)^2}{8} \right]\\ 
\label{eq:25d}
& =  1-\frac{t^2\langle\Delta E_{V_\lambda}^2\rangle}{8}\ ,
\end{align}
\ees
where to get Eq.~\eqref{eq:25b} we used the fact that the variance $\braket{X^2}-\braket{X}^2$ of the random variable $X \equiv |\bra{\psi}U_\lambda(t)\ket{\psi}|$ is positive, and to arrive at Eq.~\eqref{eq:25c} we used Eq.~\eqref{app_proof2}. This proves Eq.~\eqref{eq:F}.

\subsection{Second characterization}
First we prove Eq.~\eqref{dephas}. We start from Eq.~\eqref{eq:rho} with the initial state in the code space, i.e., 
$\rho(t) = \int d\lambda\ p(\lambda) U_{\lambda}(t) \rho(0) U^{\dag}_{\lambda}(t)$, where $\rho(0) = \PC \rho(0) \PC$ and [Eq.~\eqref{eq:Heff}] $U_\lambda(t) \PC {\rightarrow} \exp(-it V'_\lambda) \PC$ in the $\Eg\rightarrow \infty$ limit.
Diagonalizing $V'_\lambda$ we have $V'_\lambda = \lambda \PC V \PC = \lambda \sum_{m}\mu_{m}\ketbra{\mu_{m}}{\mu_{m}}$. Thus,
\bes
\begin{align}
\rho_{mn}(t)  &= \int d\lambda\ p(\lambda) \bra{\mu_m}U_{\lambda}(t) \rho(0) U^{\dag}_{\lambda}(t)\ket{\mu_n} \\
&\rightarrow \int d\lambda\ p(\lambda) \bra{\mu_m}e^{-it V'_\lambda} \rho(0) e^{it V'_\lambda}\ket{\mu_n} \\
& =\int d\lambda\ p(\lambda) e^{-it \lambda[\mu_m-\mu_n]} \bra{\mu_m}\rho(0)\ket{\mu_n} \\
& =\tilde{p}(t[\mu_{m}-\mu_{n}]) \rho_{mn}(0) \ .
\end{align}
\ees

Next we prove that $|\tilde{p}(\alpha)|  = 1-\frac{1}{2}\textrm{var}(\lambda)\alpha^2+O(\alpha^3)$ as claimed in the main text. Note that 
\beq
i^n\left.\frac{d^n\tilde{p}}{d\alpha^n}\right|_{\alpha=0} =  \int d\lambda\ p(\lambda) \lambda^n \left. e^{-i \lambda \alpha}\right|_{\alpha=0} = \langle \lambda^n \rangle\ .
\eeq
Therefore $\tilde{p}(\alpha) = 1-i \langle \lambda \rangle \alpha - \frac{1}{2} \braket{\lambda^2} \alpha^2 + O(\alpha^3)$, and 
\bes
\begin{align}
|\tilde{p}(\alpha)|  &= \left[1+ \left(\langle \lambda \rangle ^2 - \braket{\lambda^2} \right)\alpha^2 + O(\alpha^3) \right]^{1/2} \\
&= 1-\frac{1}{2}\textrm{var}(\lambda)\alpha^2+O(\alpha^3)\ .
\end{align}
\ees

\subsection{Further illustration of the IDS}
As a further illustration of the IDS consider stabilizer codes \cite{Gottesman:1996fk} in the context of quantum error suppression. Let $V$ be a logical error, i.e., a tensor product of Pauli operators which cannot be detected by a given stabilizer code $\mC$ because it exceeds the distance of the code, i.e., it commutes with the stabilizer group though it is not an element of this group. Then $\PC  V \PC \not\propto \PC $ and $\mathcal{C}$ can be decomposed into the direct sum of two subspaces, corresponding to the $\pm 1$ eigenvalues of $V$.  It follows from the definition Eq.~\eqref{eq:IDS} that $\Delta E_V=2$, meaning that the code does not provide any suppression of the errors that it cannot detect. In contrast, it completely suppresses the errors that it can detect. Non-stabilizer codes might exist that do not necessarily have high distance but do provide some suppression of high-weight errors. We will show that such codes do not exist in the setting of commuting $2$-local Hamiltonians.


\section{Proof of lemma~\ref{No-hiding}}

Recall that Lemma~\ref{lem:bound-for-2qudits} follows from Lemma~\ref{No-hiding}, so we prove the latter first.

We restate Lemma~\ref{No-hiding} here for convenience. Let $\mathcal{H}_{2}\subset \mathcal{H}_A\otimes  \mathcal{H}_B$ be an arbitrary two-dimensional subspace of $ \mathcal{H}_A\otimes  \mathcal{H}_B$. Then,
 \begin{align}
 \max_{\substack{\ket{\psi},\ket{\phi}\in \mathcal{H}_{2}\\ \langle\phi \ket{\psi}=0}} &\|\Tr_{A}\left(\ketbra{\psi}{\psi}\right)-\Tr_{A}\left(\ketbra{\phi}{\phi}\right) \|_{1}\notag \\  &+\|\Tr_{B}\left(\ketbra{\psi}{\psi}\right)-\Tr_{B}\left(\ketbra{\phi}{\phi}\right) \|_{1} \ge \frac{2}{3}
 \end{align}
where the maximization is over all pairs of orthonormal states in  $\mathcal{H}_{2}$.

\begin{proof}
Let
\begin{align}
M\equiv  \max_{\substack{\ket{\psi},\ket{\phi}\in \mathcal{H}_{2}\\ \langle\phi \ket{\psi}=0}}  &\|\Tr_{A}\left(\ketbra{\psi}{\psi}\right)-\Tr_{A}\left(\ketbra{\phi}{\phi}\right) \|_{1}\notag \\  &+\|\Tr_{B}\left(\ketbra{\psi}{\psi}\right)-\Tr_{B}\left(\ketbra{\phi}{\phi}\right) \|_{1} 
 \end{align}
Assume $\ket{0}$ and $\ket{1}$ are an arbitrary pair of orthonormal states in  $\mathcal{H}_{2}$. Let $\rho^{(A)}_{0}= \Tr_{B}\left(\ketbra{0}{0}\right)$ and $\rho^{(A)}_{1}= \Tr_{B}\left(\ketbra{1}{1}\right)$ be the reduced state of system $A$ for states  $\ket{0}$ and $\ket{1}$ respectively.  
Define
$\ket{\pm}_{s}\equiv (\ket{0}\pm i^{s}\ket{1})/{\sqrt{2}}$, where $s\in \{0,1\}$,
and let
\begin{equation} 
m\equiv \max_{s\in\{0,1\}}\|\Tr_{A}\left(\ket{+}_{s}\!\bra{+}\right)-\Tr_{A}\left(\ket{-}_{s}\!\bra{-}\right) \|_{1}\ .
\end{equation}
Then clearly 
\begin{equation}\label{maxim}
M \ge \max\{m, \|\rho^{(A)}_{0}-\rho^{(A)}_{1}\|_{1}\} \ .
\end{equation}
We proceed by finding a lower bound on $m$.

Let $X \equiv\Tr_{A}\left(\ket{0}\bra{1}\right)$. Then
\bes
\label{bound-m}
\begin{align}
m &=\max_{s\in\{0,1\}}\| (-i)^{s} X+i^{s} X^{\dag} \|_{1} \\ 
&=\max \{\|X+X^{\dag} \|_{1},  \|X-X^{\dag} \|_{1}\} \\ 
&\ge \frac{1}{2} \left( \|X+X^{\dag} \|_{1} + \|X-X^{\dag} \|_{1}\right) \\
&\ge \|X \|_{1}\ ,
\end{align}
\ees
where the last line follows from 
$X = (X+X^\dag + X -X^\dag)/2$ and 
the triangle inequality.

To determine $\|X \|_{1}$ we use Uhlman's theorem for the fidelity between two states $\rho$ and $\sigma$,
\beq
F(\rho,\sigma) \equiv \Tr\sqrt{\sqrt{\rho}\sigma\sqrt{\rho}} = \max_{\ket{\psi}} |\!\bra{\phi}\!{\psi}\rangle |\ ,
\eeq
where $\ket{\phi}$ and $\ket{\psi}$ are, respectively, a fixed and an arbitrary purification of $\sigma$ and $\rho$ \cite{nielsen2000quantum}. Since $\ket{1}$ is a fixed purification of $\rho_1^{(A)}$ and an arbitrary purification of $\rho_0^{(A)}$ can be written as $\ket{\psi} = I^{(A)}\otimes U^{(B)} \ket{0}$, where $I^{(A)}$ is the identity on system $A$ [since $\Tr_B(\ket{\psi}\!\bra{\psi}) = \Tr_B(\ket{0}\!\bra{0})$], this implies 
\begin{equation}
F(\rho^{(A)}_{0},\rho^{(A)}_{1})=\max_{U^{(B)}}\left| \bra{1} I^{(A)}\otimes U^{(B)} \ket{0} \right| \ ,
\end{equation}
where the maximization is over all unitaries which act on $\mathcal{H}_B$.  It follows that
\beq
\label{Eq:Fid}
F(\rho^{(A)}_{0},\rho^{(A)}_{1})=\max_{U^{{(B)}}}\left| \Tr(U^{(B)} X)\right|=  \|X \|_{1}\ ,
\eeq
where in the last inequality we used \cite{Bhatia:book}
\beq
\|Y\|_{1}=\max_{U: UU^\dag=I}\left|\Tr(YU)\right |\ .
\label{eq:_1}
\eeq 
This together with Eq.~\eqref{bound-m} implies
that $m\ge F(\rho^{(A)}_{0},\rho^{(A)}_{1})$, so that using Eq.~\eqref{maxim} we find
\begin{equation}
M\ge \max\{F(\rho^{(A)}_{0},\rho^{(A)}_{1}),\|\rho^{(A)}_{0}-\rho^{(A)}_{1})\|_{1} \} \ .
\end{equation}
Then, using the inequality \cite{Fuchs:99}
\begin{equation}
1-D(\rho^{(A)}_{0},\rho^{(A)}_{1}) \le F(\rho^{(A)}_{0},\rho^{(A)}_{1})\ ,
\end{equation}
where $D(\rho,\sigma) \equiv \frac{1}{2}\|\rho-\sigma\|_{1}$ is the trace-norm distance, which satisfies $0\leq D \leq 1$, we find
\begin{align}
M \ge  \max\{1-D(\rho^{(A)}_{0},\rho^{(A)}_{1}), 2D(\rho^{(A)}_{0},\rho^{(A)}_{1})\}\ge \frac{2}{3}\ .
\end{align}
\end{proof}


\section{Proof of lemma \ref{lem:bound-for-2qudits}}
Let us restate lemma \ref{lem:bound-for-2qudits}  for convenience. Let $P$ be a projector in $\mathcal{H}_A\otimes  \mathcal{H}_B$ with rank larger than one. Then there exists a single-site operator $X$ such that for any complex number $\alpha$
 \begin{equation}
  \| P X P-\alpha P\| \ge \frac{\|X\|}{6}\ .
\label{eq:bound-for-2qudits-2}
\end{equation}

\begin{proof}
By assumption $P$ has rank at least two, so its support (the linear space orthogonal to its kernel) has 
at least two orthonormal elements 
spanning a two-dimensional subspace $\mathcal{H}_{2} \subseteq\textrm{supp}(P)$. 
It follows that
\bes
\begin{align}
& \|P(c_{A} X_A +c_{B} X_B) P-\alpha P\| \notag \\
&= \|P(c_{A} X_A +c_{B} X_B-\alpha \openone) P\| \notag \\ 
\label{eq:40a}
&  \ge \max_{\ket{i}\in\textrm{supp}(P)} |\bra{i}  (c_{A} X_A +c_{B} X_B-\alpha\openone)\ket{i} |\\ 
\label{eq:41b}
&\ge \frac{1}{2}\sum_{i\in\{\psi,\phi\}}|\bra{i}(c_{A} X_A +c_{B} X_B-\alpha\openone)\ket{i} | \ ,
\end{align}
\ees
where in line~\eqref{eq:40a} we used the definition of the operator norm and in line~\eqref{eq:41b} $\ket{\psi}$ and $\ket{\phi}$ are two arbitrary pair of orthonormal states in $\mathcal{H}_{2}$. Applying the triangle inequality we then obtain
\bes
\begin{align}
& \|P(c_{A} X_A +c_{B} X_B) P-\alpha P\| \\
&\ge \frac{1}{2}[\left|\bra{\psi}  (c_{A} X_A +c_{B} X_B-\alpha\openone)\ket{\psi} \right. \notag \\
& \qquad \qquad \left. -\bra{\phi}  (c_{A} X_A +c_{B} X_B-\alpha\openone)\ket{\phi} \right|]\\ 
&=  \frac{1}{2} |\Tr \left(\left[\Tr_{A}(\ketbra{\psi}{\psi})-\Tr_{A}(\ketbra{\phi}{\phi}) \right] c_{B} X_B\right) \notag \\ 
&\qquad\qquad  +\Tr \left(\left[\Tr_{B}(\ketbra{\psi}{\psi})-\Tr_{B}(\ketbra{\phi}{\phi}) \right] c_{A} X_A\right) |
\end{align}
\ees
By multiplying $X_A$ (or $X_B$) by $-1$ we can always ensure that $\Tr \left(\left[\Tr_{A}(\ketbra{\psi}{\psi})-\Tr_{A}(\ketbra{\phi}{\phi}) \right] c_{B} X_B\right)$ and $\Tr \left(\left[\Tr_{B}(\ketbra{\psi}{\psi})-\Tr_{B}(\ketbra{\phi}{\phi}) \right]  c_{A} X_A\right)$ have the same sign. Therefore,
\bes
\begin{align}
&\max_{\substack{X_A,X_B\\ \|X_A\|=\|X_B\|=1}}\|P(c_{A} X_A+c_{B} X_B) P-\alpha P\| \\  
&\ge \frac{1}{2} \max_{\substack{X_A,X_B\\ \|X_A\|=\|X_B\|=1}} |\Tr \left(\left[\Tr_{A}(\ketbra{\psi}{\psi}-\ketbra{\phi}{\phi}) \right] c_{B} X_B\right) \notag \\ 
& \qquad\qquad  +\Tr \left(\left[\Tr_{B}(\ketbra{\psi}{\psi}-\ketbra{\phi}{\phi}) \right] c_{A} X_A\right) | \\&= \frac{|c_{B}|}{2} \max_{\substack{X_B\\ \|X_B\|=1}}  |\Tr \left(\left[\Tr_{A}(\ketbra{\psi}{\psi}-\ketbra{\phi}{\phi}) \right] X_B\right)|  \\ 
&\qquad\qquad   +\frac{|c_{A}|}{2}\max_{\substack{X_A\\ \|X_A\|=1}}  |\Tr \left(\left[\Tr_{B}(\ketbra{\psi}{\psi}-\ketbra{\phi}{\phi}) \right] X_A\right)| \notag \\ 
&= \frac{1}{2}[|c_{B}|\ \|\Tr_{A}\left(\ketbra{\psi}{\psi}-\ketbra{\phi}{\phi}\right) \|_{1}+\notag \\
&\qquad \qquad |c_{A}|\ \|\Tr_{B}\left(\ketbra{\psi}{\psi}-\ketbra{\phi}{\phi}\right) \|_{1}]\ ,
\label{eq:41d}
\end{align}
\ees
where the last equality follows from Eq.~\eqref{eq:_1}. 
Next, we note that since the result so far holds for any pair of orthonormal states $\{\ket{\psi},\ket{\phi}\} \in \mathcal{H}_{2} \subseteq\textrm{supp}(P)$, it holds in particular for the orthonormal pair that maximizes the expression in line~\eqref{eq:41d}, i.e., 
\bes\label{eq:47}
\begin{align}
&\max_{\substack{X_A,X_B\\ \|X_A\|=\|X_B\|=1}}\|P(c_{A} X_A+c_{B} X_B) P-\alpha P\| \\ 
&\ge \frac{1}{2}\max_{\substack{\ket{\psi},\ket{\phi}\in \mathcal{H}_2 \\ \<\psi\ket{\phi}=0 }} \left[ |c_{B}|\ \|\Tr_{A}\left(\ketbra{\psi}{\psi}-\ketbra{\phi}{\phi}\right) \|_{1} \right.\notag \\ 
 &\qquad\qquad  \left. + |c_{A}|\ \|\Tr_{B}\left(\ketbra{\psi}{\psi}-\ketbra{\phi}{\phi}\right) \|_{1} \right] \\ 
&\ge \frac{1}{2}\max_{\substack{\ket{\psi},\ket{\phi}\in \mathcal{H}_2 \\ \<\psi\ket{\phi}=0 }} \min\{|c_{A}|,|c_{B}|\}\left[ \|\Tr_{A}\left(\ketbra{\psi}{\psi}-\ketbra{\phi}{\phi}\right) \|_{1} \right.\notag \\ 
 &\qquad\qquad  \left. +  \|\Tr_{B}\left(\ketbra{\psi}{\psi}-\ketbra{\phi}{\phi}\right) \|_{1} \right] \\ 
 &\ge \frac{\min\{|c_{A}|,|c_{B}|\}}{3} \ ,
\end{align}
\ees
where the last inequality follows from Lemma~\ref{No-hiding}.

Now note that if $X$ only acts on either $A$ or $B$ then
\begin{align}
&\max_{\substack{X\\ \|X\|=1}}   \| P X P-\alpha P\| \notag \\
&\qquad   \geq \max_{\substack{X_A,X_B\\ \|X_A\|=\|X_B\|=1}}   \| P \frac{X_A+X_B}{2}  P-\alpha P\|\ , \end{align}
since $X_A$ acts only $A$ and $X_B$ acts only on $B$. Combining this with Eq.~\eqref{eq:47} while setting $c_A = c_B = 1/2$ yields 
\beq
\max_{\substack{X\\ \|X\|=1}}   \| P X P-\alpha P\|  \ge \frac{1}{6}\ ,
\eeq 
which implies that there exists an operator $X$ such that 
\begin{equation}
  \| P X P-\alpha P\| \ge \frac{\|X\|}{6} \ ,
 \end{equation}
 which is the claim of Lemma~\ref{lem:bound-for-2qudits}.
\end{proof}


\section{Proof of Lemma~\ref{lemma:main}}
We now focus on Hamiltonians which are sums of two-body commuting terms.  
It turns out the algebras generated by two-body commuting terms have a very simple structure. Consider a set of commuting two-body observables $\{H^{(i,j)}=H^{(j,i)}\}$ where $H^{(i,j)}$ acts nontrivially only on sites $i$ and $j$.  Then, for any site $i$ we can find a complete set of 
orthogonal projectors $\{\Pi_{\mu}^{(i)}\}$ with the following properties: i) All terms $H^{(i,j)}$ are block-diagonal with respect to these projectors, i.e., $\Pi_{\mu}^{(i)}H^{(i,j)}=H^{(i,j)} \Pi_{\mu}^{(i)}=\Pi_{\mu}^{(i)}H^{(i,j)} \Pi_{\mu}^{(i)} $ for all $H^{(i,j)}$, and  ii) For $j\neq k$ two operators $\Pi_{\mu}^{(i)} H^{(i,j)} \Pi_{\mu}^{(i)} $ and  $\Pi_{\mu}^{(i)} H^{(i,k)} \Pi_{\mu}^{(i)} $ act on different non-overlapping virtual subsystems \cite{Zanardi:2001b} of the support of $\Pi_{\mu}^{(i)}$. In other words, for any site $i$, there exists a subsystem decomposition of its Hilbert space $\mathcal{H}^{(i)}$ as 
\begin{equation}\label{decom1}
\mathcal{H}^{(i)}\cong\bigoplus_{\mu} \mathcal{H}^{(i)}_{\mu}\quad ; \qquad \mathcal{H}^{(i)}_{\mu}\cong \mathbb{C}^{d_{\mu}}\bigotimes_{j\neq i}  \mathcal{H}^{(i,j,\mu)} 
\end{equation} 
such that $\Pi^{(i)}_{\mu}H^{(i,j)}\Pi^{(i)}_{\mu}$ acts trivially on all the subsystems except $\mathcal{H}^{(i,j,\mu_{i})}\otimes \mathcal{H}^{(j,i,\mu_{j})}$. Here $\mathbb{C}^{d_{\mu}}$ is a multiplicity subsystem on which all $\{H^{(i,j)}\}$ act trivially.
The proof is a straightforward application of the decomposition of $C^{\ast}$-algebras into irreducible matrix algebras \cite{Landsman:98a}. 

Since the projector onto the ground subspace $\PC$ is in the algebra generated by $\{H^{(i,j)}\}$,\footnote{In general, for any Hamiltonian which is sum of terms,  the projector onto the ground subspace (like projectors onto other eigen-subspaces of Hamiltonian) is in the algebra generated by these terms. This can be easily seen, for instance, by noting that  the projector onto the ground subspace is a function of the Hamiltonian. If all the local terms commute, then this algebra is commutative, and therefore the projector onto the ground subspace also commutes with all the local terms.} it follows that $\PC$ has also the same block-diagonal structure, i.e., 
\begin{equation}\label{commutation-proj}
\PC \bigotimes _{i} \Pi_{\mu_{i}}^{(i)}=\bigotimes _{i} \Pi_{\mu}^{(i)} \PC= \bigotimes_{(i,j)} P_{\mu_{i},\mu_{j}}^{(i,j)}
\end{equation} 
where $P_{\mu_{i},\mu_{j}}^{(i,j)}$ is a projector which acts trivially on all subsystems except $ \mathcal{H}^{(i,j,\mu_{i})}\otimes \mathcal{H}^{(j,i,\mu_{j})}$.  

In general, for any site $i$ the projector onto the ground subspace $\PC$ can have support in more than one sector $\mu$. In this case, since $\PC$ commutes with all $\{\Pi_{\mu}^{(i)}\}$ it follows that there exists ground states  $|\psi_{i}^{\mu}\rangle$ and $|\psi_{i}^{\nu}\rangle$ for different sectors $\mu$ and $\nu$ such that  $\Pi_{\mu}^{(i)} |\psi_{i}^{\mu}\rangle=|\psi_{i}^{\mu}\rangle $ and $\Pi_{\nu}^{(i)}|\psi_{i}^{\nu}\rangle=|\psi_{i}^{\nu}\rangle $. Furthermore, any superposition $(|\psi_{i}^{\mu}\rangle+e^{i\theta}|\psi_{i}^{\nu}\rangle)/\sqrt{2}$ of these states is also in the ground subspace. However, the different ground states for different values of $\theta$ can be transformed into each other by single-site errors: the single-site unitary $\exp(i\pi \Pi_{\nu}^{i})$ transforms the ground state  $(|\psi_{i}^{\mu}\rangle+e^{i\theta}|\psi_{i}^{\nu}\rangle)/\sqrt{2}$ into the orthogonal ground state $(|\psi_{i}^{\mu}\rangle+e^{i(\theta+\pi)}|\psi_{i}^{\nu}\rangle)/\sqrt{2}$. This means that if $\PC$ has support in more than one sector there is no protection whatsoever against single-site errors, and so Theorem~\ref{Thm:main} trivially holds by choosing the perturbation as ${X}=\exp(i\pi \Pi_{\nu}^{i})$ in Eq.~\eqref{eq:1/3}. Hence, to prove Theorem~\ref{Thm:main} we assume from now on that for each site $i$, the projector onto the ground subspace $\PC$ has support in only one sector $\mu_{i}$. Therefore $\PC$ has the following form
\begin{equation}\label{proj}
\PC=\bigotimes_{(i,j)} P_{\mu_{i},\mu_{j}}^{(i,j)}
\end{equation}
This means that $\PC$ is the tensor product of the projectors onto the ground subspace of a non-interacting set of two-body systems. In other words, the protections that two-body commuting Hamiltonians can provide against local noise is limited by the protection we can obtain by encoding quantum information in a system formed from only two local sites, and as we have seen before, this protection is limited by  the no-hiding theorem.  

To see this more clearly assume  that there is a  pair of sites $i$ and $j$ for which $P_{\mu_{i},\mu_{j}}^{(i,j)}$ has rank larger than one. Then from Lemma \ref{lem:bound-for-2qudits} we know that there exists a single-site operator  $Y^{(i)}$ which acts non-trivially only on the subsystem $\mathcal{H}^{(i,j,\mu_{i})}$ such that for any value of $\alpha$, 
\beq
\| P_{\mu_{i},\mu_{j}}^{(i,j)} Y^{(i)}P_{\mu_{i},\mu_{j}}^{(i,j)}-\alpha P_{\mu_{i},\mu_{j}}^{(i,j)}\|\ge \frac{\|Y^{(i)}\|}{6}\ .
\label{eq:Yi}
\eeq 
Since $Y^{(i)}$ commutes with all projectors $\{P_{\mu_{k},\mu_{l}}^{(k,l)}\}$ except $P_{\mu_{i},\mu_{j}}^{(i,j)}$, using Eq.~\eqref{proj} we find that for any $\alpha$
\begin{align}
\|\PC&Y^{(i)}\PC-\alpha \PC\| \\ 
&=\|\bigotimes_{(k,l)\neq(i,j) } P_{\mu_{k},\mu_{l}}^{(k,l)}\otimes  [P_{\mu_{i},\mu_{j}}^{(i,j)} Y^{(i)}P_{\mu_{i},\mu_{j}}^{(i,j)} 
-\alpha P_{\mu_{i},\mu_{j}}^{(i,j)}] \| \notag \\ 
&=\|P_{\mu_{i},\mu_{j}}^{(i,j)} Y^{(i)} P_{\mu_{i},\mu_{j}}^{(i,j)}-\alpha P_{\mu_{i},\mu_{j}}^{(i,j)}\| \ .
\end{align}
Combining this with Eq.~\eqref{eq:Yi} proves Lemma~\ref{lemma:main}. 


\end{document}